\documentclass[prb,nofootinbib,twocolumn,showpacs,floatfix]{revtex4-1}
\pdfoutput=1
\usepackage{bm}
\usepackage{amsfonts}
\usepackage[pdftex]{color,graphicx}
\usepackage{amssymb}
\usepackage{amsmath}
\usepackage{bbm}
\usepackage{amsthm}
\usepackage{hyperref}
\hypersetup{
        colorlinks=true,
}

\usepackage{times}
\usepackage[english]{babel}
\usepackage{blindtext}

\newcommand{\tbf}[1]{\textbf{#1}}

\newcommand{\donehalf}{\dfrac{1}{2}}
\newcommand{\onehalf}{\frac{1}{2}}
\newcommand{\infint}[1]{\int\limits_{-\infty}^{\infty}{#1}}
\newcommand{\tr}{\text{Tr}}
\newcommand{\meas}[2]{\text{d}^\ensuremath{{#1}\text{#2}}\,}
\newcommand{\abs}[1]{\left| #1 \right|} 

\newcommand{\ket}[1]{\left| #1 \right>}
\newcommand{\bra}[1]{\left< #1 \right|}
\newcommand{\braket}[2]{\left< #1 | #2 \right>}

\newcommand{\mc}[1]{\ensuremath{\mathcal{#1}}}

\newcommand{\dtau}{\Delta\tau}
\newcommand{\expv}[1]{\left\langle #1 \right\rangle}

\newcommand{\Renyi}{R\'{e}nyi\ }

\let\oldmarginpar\marginpar
\renewcommand\marginpar[1]{\-\oldmarginpar[\raggedleft\tiny\color{red} #1]%
{\raggedright\tiny #1}}

\graphicspath{{Figures/}}

\begin{document}

\title{\Renyi Entropies of Interacting Fermions from Determinantal Quantum Monte Carlo Simulations}
\date{\today}

\author{Peter Broecker}
\author{Simon Trebst}
\affiliation{Institute for Theoretical Physics, University of Cologne, 50937 Cologne, Germany}


\begin{abstract}
Entanglement measures such as the entanglement entropy have become an indispensable tool to identify the fundamental character of ground states of interacting quantum many-body systems. For systems of interacting spin or bosonic degrees of freedom much recent progress has been made not only in the analytical description of their respective entanglement entropies but also in their numerical classification. 
Systems of interacting fermionic degrees of freedom, however, have proved to be more difficult to control, in particular with regard to the numerical understanding of their entanglement properties.
Here we report a generalization of the replica technique for the calculation of \Renyi entropies to the framework of 
determinantal Quantum Monte Carlo simulations -- the numerical method of choice for unbiased, large-scale simulations of interacting fermionic systems. 
We demonstrate the strength of this approach over a recent alternative proposal based on a decomposition in free fermion Green's functions by studying the entanglement entropy of one-dimensional Hubbard systems both at zero and finite temperatures.
\end{abstract}

\pacs{05.30.-d, 02.70.Ss, 03.67.Mn, 89.70.Cf}

\maketitle


\section{Introduction}
Entanglement is one of the fundamental concepts of quantum mechanics that describes how quantum mechanical objects -- e.g. photons, electrons, or spins -- are interwoven into a collective state~\cite{EPR}. 
If such a state can no longer be described as a simple product state of single-particle wave functions, one says that the quantum mechanical objects are entangled. 
Beyond its conceptual relevance quantum mechanical entanglement has turned into a key resource in various fields of modern physics~\cite{Eisert}. 
In quantum information theory it is exploited in storing and manipulating information in so-called qubits \cite{Shannon,vonNeumann,Wehrl}. 
In condensed matter physics entanglement has become increasingly appreciated as a measure to classify different states of quantum matter which cannot be distinguished by any local observable such as topologically ordered states \cite{Wolf,Wen90,KitaevPreskill,LevinWen}. 
The probably stunning realization that oftentimes ground states of quantum many-body systems are far from being highly entangled states has led to the development of a novel class of (tensor network) algorithms to simulate quantum many-body systems in a variational low-entanglement approach \cite{DMRG,PEPS,MERA}. 

\begin{figure}[b]
  \centering
  \includegraphics[width=.75\columnwidth]{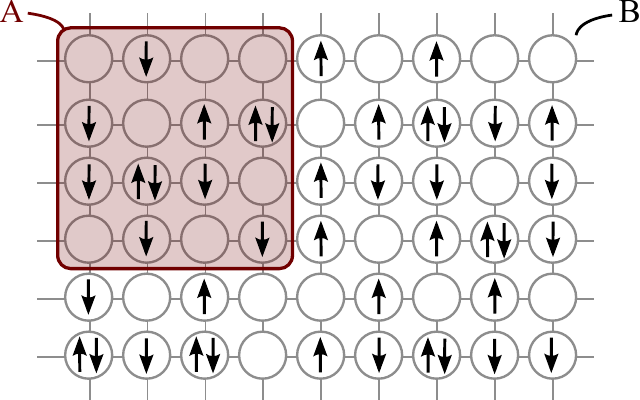}
    \caption{The bipartition of a quantum many-body system into part $A$ and its complement $B$.
    		\label{fig:bipartition}}
\end{figure}

While the notion of entanglement was originally associated with typically a handful of qubits its application to quantum many-body systems requires entanglement measures that allow to deal with an almost arbitrarily large number of interwoven quantum mechanical degrees of freedom. 
One such powerful measure is the so-called entanglement entropy~\cite{Eisert}, which can be calculated from a bipartition of a quantum many-body system into 
two complimentary parts $A$ and $B$ as illustrated in Fig.~\ref{fig:bipartition}. 
Tracing out the degrees of freedom in one subsystem one can calculate a reduced density matrix for the other, e.g. $\rho_A = {\rm Tr}_B(\ket{\psi}\bra{\psi})$. 
The information in the density matrix is then condensed into a single number, e.g. the von Neumann entropy~\cite{vonNeumann}
\begin{equation}
     S(A)=-\tr \,[\rho_A \log \rho_A] \,.
     \label{eq:EntanglementEntropy}
\end{equation}
The von Neumann entropy is the most prominent member of a more general family of entanglement entropies, the so-called \Renyi entropies
\cite{Renyi} which are calculated from the density matrix as
\begin{equation}
     S_n(A) = \dfrac{1}{1 - n} \log{\left(\, \tr(\rho_A^n) \,\right)} \,,
     \label{eq:RenyiEntropy}
\end{equation}
where the limit $n\rightarrow 1$ recovers the above von-Neumann entropy.
The strength of these entropic entanglement measures becomes apparent when considering the scaling of the entanglement entropy for varying sizes of the subsystem $A$. 
In contrast to conventional thermodynamic entropies the entanglement entropy is not extensive, but rather scales with the length of the boundary between the two partitions -- the so-called boundary law~\cite{Wolf} (which is often also referred to as area-law). 
Corrections to this prevalent boundary law have received widespread attention for their ability to classify different states of quantum matter~\cite{Eisert}. 
For instance, it has been shown that the topological character of non-local order present in a quantum ground-state wavefunction can be revealed by a universal $O(1)$ correction \cite{KitaevPreskill,LevinWen} to the boundary law. 
Numerical simulations of quantum spin systems have subsequently shown that such an identification is indeed feasible and unambiguously revealed the long conjectured topological order present in ground states of certain frustrated quantum magnets \cite{Jiang12a,Jiang12b,Schollwoeck}.
Systems of interacting {\em fermionic} degrees of freedom are the conceptually most interesting class of interacting quantum many-body systems due to the even more complex variety of possible ground states and corresponding entropic signatures arising from the intricate nodal structure of fermionic wave functions. While a generalization of the numerical schemes to calculate entanglement entropies also for these many-fermion systems is highly desirable, progress so far has been limited to variational Monte Carlo techniques~\cite{Zhang,McMinis,swingle_oscillating_2013,rezayi_entanglement_2014}. The first step to develop an approach based on determinantal Monte Carlo -- an unbiased, auxiliary field technique that has become the method of choice for large-scale simulations of interacting fermion systems -- has recently been reported by Grover~\cite{grover_entanglement_2013}, whose approach is based on a decomposition of the entanglement entropy in terms of free-fermion Green's functions.

The purpose of this manuscript is to describe an efficient numerical method to calculate the entanglement entropy for systems of interacting fermions in any spatial dimension. 
Our approach generalizes a replica scheme used to calculate \Renyi entropies in world line quantum Monte Carlo approaches for interacting spin or bosonic systems to the framework of determinantal Monte Carlo simulations as detailed in the following section.

We demonstrate the applicability of this approach by simulating one-dimensional Hubbard systems and discuss the strength of our technique in a detailed comparison with Grover's recent alternative proposal to decompose the entanglement entropy in terms of free-fermion Green's functions~\cite{grover_entanglement_2013} in Section \ref{sec:Hubbard}. 
We close with an outlook in Sec.~\ref{sec:outlook}.


\section{Determinantal QMC and the replica trick}
\label{sec:DQMC-ReplicaTrick}

We consider a setup where the interactions between spinful fermionic degrees of freedom are  captured by a lattice Hamiltonian such as the Hubbard model
\begin{eqnarray}
	\mc{H} = -t \sum\limits_{\langle i,j\rangle,\sigma}\left({c_{i,\sigma}^\dagger c_{j,\sigma}^{\phantom\dagger}}+{\rm h.c.}\right) 
		      +\, U \sum\limits_{i} n_{i,\uparrow} n_{i,\downarrow} \nonumber \\ 
		      - \mu \sum\limits_{i} \left( n_{i,\uparrow} + n_{i,\downarrow}\right) \,,
\end{eqnarray}
whose physics thrives from the competition of the on-site interaction $U$ and the hopping $t$ for fixed chemical potential $\mu$.
Our method is, however, not limited to the specifics of Hubbard model but can in fact be applied to any fermionic Hamiltonian amenable to a quantum Monte Carlo simulation.
The quantum mechanical state of such an interacting many-fermion system can be described via its density matrix $\rho$.
When considering the ground state of the system at zero temperature this density matrix is given by
\begin{equation}
  \rho = \dfrac{\ket{\psi}\bra{\psi}}{\braket{\psi}{\psi}} \,,
  \label{eq:rho_rewritten}
\end{equation}
while at finite temperatures it takes the form
\begin{equation}
	\rho = \dfrac{ \exp{\left( -\beta \mc{H} \right)}}{\tr( \exp{\left( -\beta \mc{H} \right)} )} \,. 
\end{equation}
In both cases, we have introduced an explicit normalization constant $\mathcal{N}$ in the denominator, which not only ensures that the trace of the so-defined density matrix is $1$, but will play an important conceptual role in the following.
To be even more explicit, we can rewrite both expressions as 
\begin{equation}\label{eq:rho_prime}
\rho = \dfrac{1}{\mc{N}}\rho^\prime \,,
\end{equation}
which is the form we will be using in the following.


\subsection{\Renyi entropies and the replica trick}

\begin{figure}[t]
\centering
   \includegraphics[width=\columnwidth]{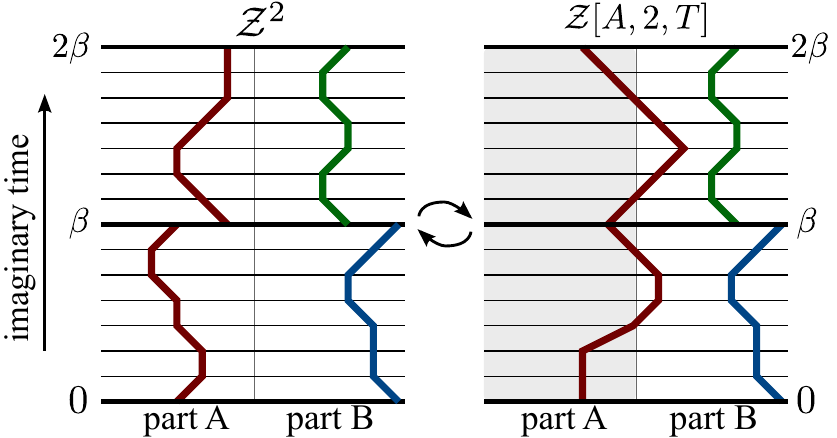}
    \caption{(color online) Ensemble switching in a world line picture. The left side shows the square of the regular partition sum $\mc{Z}^2$ 
    		 where all world lines have to be $\beta$-periodic. 
		 The right side shows a configuration of allowed world lines in the $\mc{Z}[A, 2, T]$ partition sum, where $2\beta$-periodicity 
		 is enforced in part $A$ of the system but part $B$ remains $\beta$-periodic.
      \label{fig:EnsembleSwitching}}
\end{figure}

The first step in calculating the \Renyi entropy is to determine the reduced density matrix $\rho_A$ by tracing out the degrees of freedom in subsystem $B$ 
\[ \rho_A = \tr_B \rho \,. \]
In our numerical calculations we will concentrate on the second \Renyi entropy, i.e. the case of $n=2$,  which can be calculated from the reduced density matrix $\rho_A$ as
\begin{equation}\label{eq:renyi_a}
S_2(A) = - \log{\left( \tr_A \left(\rho_A^2 \right) \right)} \,.
\end{equation}
Note that $\rho_A$ is squared before the remaining degrees of freedom of subsystem $A$ are traced out. 
Using the notation of Eq.~\eqref{eq:rho_prime}, we thus find the following  general expression for the \Renyi entropy%
\begin{equation}
S_2(A) = - \log{\left( \dfrac{\tr_A \left(\rho_A^{\prime\, 2} \right)}{\mc{N}^2} \right) } \,,
  \label{eq:S2}
\end{equation}
which is valid for both finite temperature and ground state considerations.
In the following we will closely examine how this definition of the \Renyi entropy~\eqref{eq:S2} can be translated into an algorithm for its numerical computation. For concreteness we will initially
focus on the finite-temperature scenario and expand our discussion to ground-state calculations in a later step.
Considering first the denominator in Eq.~\eqref{eq:S2}, we note that
\begin{equation}
\mc{N}^2 = \left(\tr{\rho}\right)^2  = \mc{Z}^2,
\end{equation}
i.e. the normalization $\mc{N}^2$ is equal to the square of the usual partition sum considered in thermodynamics. 
The numerator of definition \eqref{eq:S2} is a bit more involved
\begin{align}\label{eq:z_num}
  \tr_A \left(\rho_A^{\prime 2} \right) &= \sum\limits_{\mathcal{A}, \mathcal{A}^\prime, \mathcal{B}, \mathcal{B}^\prime}
  \bra{\mathcal{A}{\mathcal{B}^\prime}}
  \rho^\prime
  \ket{\mathcal{A}^\prime{\mathcal{B}^\prime}}
  \bra{\mathcal{A}^\prime{\mathcal{B}}}
  \rho^\prime
  \ket{\mathcal{A}\mathcal{B}}\nonumber \\
  &\equiv \mc{Z}[A, 2, T] \,,
\end{align}
where we have defined the partition function $\mc{Z}[A, 2, T]$.
We are now left with the question of how to numerically calculate these two partition functions. 
To this end, we will first consider their calculation in the framework of world line quantum Monte Carlo techniques, as they are typically used for systems of interacting spin or bosonic degrees of freedom. 
We will then turn to the framework of determinantal quantum Monte Carlo (DQMC) techniques, typically used to simulate many-fermion systems.

Turning first to the case of world line QMC techniques, it is helpful to translate the two partition sums $\mc{Z}^2$ and $\mc{Z}[A, 2, T]$ into their respective world line representations as illustrated in Fig.~\ref{fig:EnsembleSwitching}.
On the left side, the world line representation of $\mc{Z}^2$  from the denominator in Eq.~\eqref{eq:S2} is depicted with two sets of $\beta$-periodic world lines extending from $0$ to $\beta$ and from $\beta$ to $2\beta$, respectively.
On the right hand site a world line representation is depicted for the partition function  $\mc{Z}[A,2,T]$ of the numerator of Eq.~\eqref{eq:S2},
where we consider two replicas of the system connected in imaginary time. Carefully implementing the imaginary time boundary conditions defined in Eq.~\eqref{eq:z_num} results in a $\beta$ periodicity for part $B$ and a $2\beta$ periodicity for part $A$ -- a scheme often referred to as the replica trick \cite{CalabreseCardy09,hastings_measuring_2010}.
To sample world line configurations according to these two partition sums, one simply needs to implement their respective imaginary time boundary conditions -- a task, which is relatively straightforward for any existing world line Monte Carlo implementation \cite{hastings_measuring_2010,melko_finite_2010}. 

\begin{figure}[t]
\centering
\includegraphics{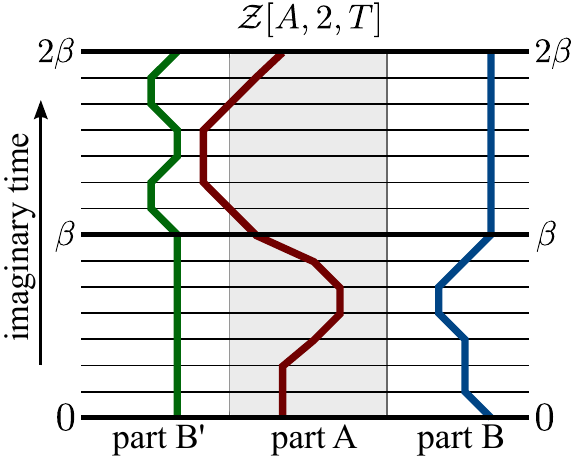}
    \caption{(color online) The enlarged simulation cell used to port the replica trick to DQMC simulations.
      \label{fig:EnsembleSwitchingUnfolded}}
\end{figure}

\subsection{Determinantal QMC}

Let us now turn to the framework of DQMC simulations and try to adapt the evaluation of the two partition functions $\mc{Z}^2$ and $\mc{Z}[A, 2, T]$ needed to calculate the \Renyi entropy of Eq.~\eqref{eq:S2} for a fermionic system. 
Describing the DQMC framework in a nutshell, one first decouples quartic terms in the Hamiltonian by a Hubbard-Stratonovich transformation introducing an auxiliary field, which then allows to integrate out the fermions analytically. A Monte Carlo scheme is then used to  sample configurations of the auxiliary field.
One of the key differences to world line techniques described above  is that we do not sample world lines directly.
In fact, by integrating out the fermionic degrees of freedom, we sample {\em all} possible world line configurations simultaneously for a given configuration of the auxiliary field. 

This raises the question of how to properly implement the replica scheme in this framework. We will concentrate our discussion on the key conceptual steps in the following and refer the inclined reader interested in a more concise technical description to appendix \ref{app:DQMC}.
Considering at first an arbitrary pair of a fermion state $\psi$ and an auxiliary field state $\sigma$ its statistical weight $W(\sigma, \psi)$
 is given by a Slater determinant
\begin{equation}
	\label{eq:fermionweight_main}
	W(\sigma, \psi) = \det\left(\sigma, \psi \right) \,.
\end{equation}
A crucial step is to realize that the grand-canonical trace over these Slater determinants can be recast as a {\em single} determinant
\begin{equation}
  \label{eq:grandcanonical_trace}
  \tr{\, \det\left(\sigma, \psi \right) } = \det{ \left(\sigma\right) } \,,
\end{equation}
which implies that one can integrate out the fermionic degrees of freedom and instead consider only configurations of the auxiliary field.
However, it should be noted that this identity works only if the weights of the partition sum can be written in terms of single Slater determinants.
When considering the replica scheme for the calculation of \Renyi entropies, however, one typically encounters statistical weights in the expression for the partition function $\mc{Z}[A, 2, T]$ that are the product of {\em two}  determinants%
\begin{align}
   &\mc{Z}[A, 2, T] =
   \sum\limits_{\{\sigma\}} \sum\limits_{\mathcal{A}, \mathcal{A}^\prime, \mathcal{B}, \mathcal{B}^\prime}
   \bra{\mathcal{A}{\mathcal{B}^\prime}}
   \rho^\prime
   \ket{\mathcal{A}^\prime{\mathcal{B}^\prime}}
   \bra{\mathcal{A}^\prime{\mathcal{B}}}
   \rho^\prime
   \ket{\mathcal{A}\mathcal{B}}\nonumber\\
   &=  \sum\limits_{\{\sigma\}} \sum\limits_{\mathcal{A}, \mathcal{A}^\prime, \mathcal{B}, \mathcal{B}^\prime}
   \det{}{\left(\sigma,\mc{A}\mc{B}^\prime\vert\mc{A}^\prime\mc{B}^\prime\right)}
   \det{}{\left(\sigma, \mc{A}^\prime\mc{B}\vert\mc{A}\mc{B}\right)} \,,\label{eq:weight_as_prod_of_dets}
\end{align}
where the two determinants are denoted as $\det{}{\left(\sigma,\mc{A}\mc{B}^\prime\vert\mc{A}^\prime\mc{B}^\prime\right)}$ and  $\det{}{\left(\sigma, \mc{A}^\prime\mc{B}\vert\mc{A}\mc{B}\right)}$. They again depend on the auxiliary field configuration $\sigma$ and the arguments $\mc{A}, \mc{B}, \mc{A}^\prime, \mc{B}^\prime$ indicate the imaginary time boundary conditions for a particular pair of auxiliary field and fermionic configuration.
This structure seems to suggest that one can no longer integrate out the fermions by taking a grand-canonical trace as in Eq.~\eqref{eq:grandcanonical_trace} -- the essential step at the heart of the DQMC framework, which if missing does not allow numerical simulations of the fermionic system at feasible numerical cost.

The key idea in our approach is to represent the replica scheme in a setting where the statistical weights can again be simplified to a form relying on a single Slater determinant thus allowing to take a grand-canonical trace of the form~\eqref{eq:grandcanonical_trace}.
This is achieved by artificially enlarging the system by considering an additional copy $B^\prime$ of subsystem $B$, which we will use to selectively evolve subsystem $B$ in imaginary time.  
In particular we will show that an imaginary-time Hamiltonian of the form
\begin{equation}\label{eq:img_time_ham}
\widetilde{\mathcal{H}}(\tau) = \mathcal{H}_{AB}\;\Theta(\tau)\;\Theta(\beta - \tau) + \mathcal{H}_{AB^\prime}\;\Theta(\tau-\beta)\;\Theta(2\beta - \tau)
\end{equation}
will give direct access to the partition sum $\mc{Z}[A, 2, T]$. A world line representation of this Hamiltonian is given in Fig.~\ref{fig:EnsembleSwitchingUnfolded}.
To see this, suppose that we are given \eqref{eq:img_time_ham} as a model Hamiltonian and our task is to determine its physics at an arbitrary temperature, suggestively written  as $2\beta$. 
We denote a given basis state by $\ket{\psi} = \ket{\mc{A}, \mc{B}, \mc{B}^\prime}$, which leads to a partition sum of
\[ \widetilde{\mc{Z}} = \sum\limits_{ \{ \mc{A}, \mc{B}, \mc{B}^\prime \} } \bra{\mc{A}, \mc{B}, \mc{B^\prime}} 
\exp{\left(-\beta\mc{H}_{AB^\prime}\right)}\exp{\left(-\beta\mc{H}_{AB}\right)} 
\ket{\mc{A}, \mc{B}, \mc{B^\prime}}.
\]
Although there are two propagation operators, the weight of the system would still be given as a {\em single} Slater determinant because only one expectation value has to be evaluated. We proceed to insert a resolution of unity in between the two exponential operators to obtain
\begin{widetext}
\begin{equation}
\widetilde{\mc{Z}} = \sum\limits_{ \{ \mc{A}, \mc{B}, \mc{B}^\prime, \mc{C}, \mc{D}, \mc{D}^\prime \} } 
\bra{\mc{A}, \mc{B}, \mc{B^\prime}} 
\exp{\left(-\beta\mc{H}_{AB^\prime}\right)}
\ket{\mc{C}, \mc{D}, \mc{D}^\prime}
\bra{\mc{C}, \mc{D}, \mc{D}^\prime}
\exp{\left(-\beta\mc{H}_{AB}\right)} 
\ket{\mc{A}, \mc{B}, \mc{B}^\prime}.
\end{equation}
\end{widetext}
Notice that in the right term, states of $\mc{B}^\prime$ do not appear in the Hamiltonian. 
Thus, independent of the specific form of $\mc{A}$ and $\mc{B}$, we need to have $\mc{B}^\prime = \mc{D}^\prime$ for any non-vanishing
term contributing to this partition function. Similarly, one obtains $\mc{B} = \mc{D}$ from inspecting the left term.
Further, if subsystems  $\mc{B}$ and $\mc{B}^\prime$ do not appear in the Hamiltonian for a given imaginary time interval they not only do not evolve, but also remain completely decoupled from the rest of the system over this interval. 
As a result, they will also not affect the statistical weight 
\begin{align*}
&\bra{\mc{C}, \mc{B}, \mc{B}^\prime}
\exp{\left(-\beta\mc{H}_{AB}\right)} 
\ket{\mc{A}, \mc{B}, \mc{B}^\prime}\\
&= \left( \bra{\mc{C}, \mc{B}} \otimes\bra{\mc{B}^\prime}\right)\,
\exp{\left(-\beta\mc{H}_{AB}\right)}\,
\left(\ket{\mc{A}, \mc{B}}\otimes \ket{\mc{B}^\prime}\right)\\
&=\bra{\mc{C}, \mc{B}}
\exp{\left(-\beta\mc{H}_{AB}\right)} 
\ket{\mc{A}, \mc{B}}
\end{align*}
and thus can safely be ignored.
Finally, renaming $\mc{C}$ to $\mc{A}^\prime$ to match our earlier notation  we  obtain the following simplified expression
\begin{widetext}
\begin{align}
\label{eq:Zfinal}
\widetilde{\mc{Z}} &= \sum\limits_{\mathcal{A}, \mathcal{A}^\prime, \mathcal{B}, \mathcal{B}^\prime}
\bra{\mathcal{A}{\mathcal{B}^\prime}}
\exp\left(-\beta\mathcal{H_{AB^\prime}}\right)
\ket{\mathcal{A}^\prime{\mathcal{B}^\prime}}
\bra{\mathcal{A}^\prime{\mathcal{B}}}
\exp\left(-\beta\mathcal{H_{AB}}\right)
\ket{\mathcal{A}\mathcal{B}} = \mc{Z}[A,2,T] \,,
\end{align}
\end{widetext}
which is precisely the expression for the sought-after partition function $\mc{Z}[A,2,T]$.
We have thus shown that one can indeed recast the partition function $\mc{Z}[A,2,T]$ in a way that relies only on single determinants thus allowing to take the grand-canonical trace \eqref{eq:grandcanonical_trace}.

\vspace{-4mm}
\subsubsection*{Ground-state formulation}
When considering the ground-state DQMC algorithm (see appendix \ref{app:GSalgorithm}) only minor modifications to the above scheme have to be implemented.
The normalization constant $\mc{N}$  introduced in \eqref{eq:rho_prime} is now given as
\[ \mc{N} = \braket{\psi}{\psi} = \sum\limits_{\ket{\mc{A}}} \braket{\psi}{\mc{A}}\braket{\mc{A}}{\psi} = \tr \left(\ket{\psi}\bra{\psi}\right). \]
The ground-state wave function $\ket{\psi}$ is obtained by a projective scheme
\begin{equation}
	 \ket{\psi} = \lim\limits_{\Theta \rightarrow \infty} e^{-\Theta \mc{H}}\ket{\psi_T} \,, 
	 \label{eq:projection_main}
\end{equation}
applied to a test wave function $\ket{\psi_T}$. If the test wave function has a non-zero overlap with the actual ground-state wavefunction,
this projective scheme should eliminate all contributions from excited states and converge to the ground-state wavefunction.
Inserting this projection into the definition of the \Renyi entropy in Eq.~\eqref{eq:S2}, we find an expression for the canonical $\tr \rho_A^{\prime\,2}$ very similar to the finite temperature expression of the grand-canonical trace for $\widetilde{\mc{Z}}$ in Eq.~\eqref{eq:Zfinal} discussed above 
\begin{widetext}
\begin{align}\label{eq:replica_gs}
\tr \,\rho_A^{\prime\,2} = \lim\limits_{\Theta \rightarrow \infty} 
\sum\limits_{\mathcal{A}, \mathcal{A}^\prime, \mathcal{B}, \mathcal{B}^\prime}
\bra{\mathcal{A}{\mathcal{B}^\prime}}
\exp\left({-\Theta\mc{H}}\right)\ket{\psi_T}\bra{\psi_T}\exp\left({-\Theta\mc{H}}\right)
\ket{\mathcal{A}^\prime{\mathcal{B}^\prime}}
\bra{\mathcal{A}^\prime{\mathcal{B}}}
\exp\left({-\Theta\mc{H}}\right)\ket{\psi_T}\bra{\psi_T}\exp\left({-\Theta\mc{H}}\right)
\ket{\mathcal{A}\mathcal{B}},
\end{align}
\end{widetext}
where the only difference is the appearance of the density matrices $\ket{\psi_T}\bra{\psi_T}$, which however come in handily when
taking the grand-canonical trace to make use of Eq.~\eqref{eq:grandcanonical_trace}. 
Precisely because of the occurrence of these density matrices only states with an occupation number identical  to the one of the test wave function will  contribute. 
\subsubsection*{Higher \Renyi entropies}
While we have concentrated our discussion on \Renyi entropies of order $2$, it should be noted that our algorithm can be extended in a straightforward way to also compute higher \Renyi entropies. For the calculation of the $n$-th \Renyi entropy via the replica trick, imaginary time has to be split into $n$ segments which would contribute one determinant each in Eq.~\eqref{eq:weight_as_prod_of_dets}. It would thus be necessary to introduce $n$ replicas of the subsystem $B$ and work in an overall system of size $N_A + n\cdot N_B$.
We have not implemented this more general case and therefore cannot comment on limiting system sizes or potential numerical instabilities arising in such an extended scheme.

\begin{figure}[t]
\centering
   \includegraphics[width=\columnwidth]{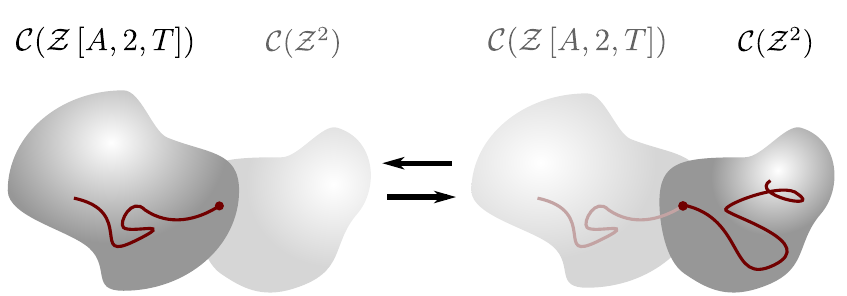}
    \caption{(color online) Schematic illustration of the ensemble switching method to calculate the \Renyi entropy of Eq.~\eqref{eq:S2}. We start a random walk in one of the configuration spaces corresponding to the ensembles appearing in the numerator and the denominator of Eq.~\eqref{eq:S2}, denoted by $\mc{C}(\mc{Z}[A, 2, T])$ and $\mc{C}(\mc{Z}^2)$, respectively. Whenever we encounter a configuration that is admissible in both ensembles, we compare their relative weights and decide in which ensemble we continue to sample configurations based on a Metropolis scheme.
      \label{fig:EnsembleSwitchingCloud}}
\end{figure}

\subsubsection*{Ensemble switching}
The replica scheme outlined above allows to directly sample partition functions of the form $\mc{Z}[A,2,T]$. 

For the calculation of the \Renyi entropy, however, we are really interested in determining the {\em ratio} of the partition functions $\mc{Z}[A,2,T]$ and $\mc{Z}^2$  as given in Eq.~\eqref{eq:S2}.
This ratio can be directly accessed~\cite{roscilde} without explicitly calculating the individual partition functions but by cleverly switching between the two ensembles.

Imagine a two state simulation, where the weight of two states is given by  $w_1$ and $w_2$.
For any simulation fulfilling detailed balance the random walk would spend $N_1 = {w_1}/{(w_1 + w_2)}$ steps in state 1 and $N_2 = {w_2}/{(w_1 + w_2)}$ steps in state 2. 
Thus, the ratio of the weights $w_1/w_2$ corresponds precisely to the relative time spent in the two respective states.
We can readily generalize~\cite{roscilde} this statement to a situation where we sample a random walk switching back and forth between two ensembles whose partition functions equal the weights $w_1 = \mc{Z}[A,2,T]$ and $w_2 = \mc{Z}^2$, respectively. Thus the ratio of relative time spent sampling each of the two ensembles can then be used to calculate the  entanglement entropy
\begin{equation}
S_2(A) = - \log{\left( \dfrac{\mc{Z}[A,2,T]}{\mc{Z}^2} \right)} =  -\log{\left( \dfrac{N_1}{N_2} \right) } \,.
\end{equation}
In practical terms, we start our simulation in one of the two ensembles and after a fixed number of Monte Carlo steps, we calculate the weight of the current configuration in both ensembles and switch ensembles according to Metropolis rules.

When implementing this ensemble switching method, one benefits from an additional advantage of the determinantal QMC framework.
The configuration space of $\mc{Z}[A,2,T]$ and $\mc{Z}^2$ is exactly equal, and the transition probabilities $p_{1\rightarrow 2}$ and $p_{2 \rightarrow 1}$ are typically spread over the entire range $(0, 1]$. 
In appendix \ref{app:ensemble_switching} we show that the Monte Carlo estimate for the \Renyi entropy is actually given as the ratio
\begin{equation}
        \label{eq:switching_ratio_main}
 	\left\langle \dfrac{N_1}{N_2} \right\rangle = \dfrac{\langle p_{2 \rightarrow 1} \rangle}{\langle p_{1\rightarrow 2}\rangle } \,,
\end{equation}
which can be obtained by two separate simulations. 
This allows for much quicker convergence than we would obtain by explicitely switching between ensembles and counting how much time we spent in each one of the two ensembles.

It should further be noted that our approach does not necessarily require to iteratively build up the subsystem $A$  from smaller blocks to achieve convergence~\cite{hastings_measuring_2010}, as it has been observed in the context of world line Monte Carlo approaches 
where the overlap between the ensembles might become rather small. We note that such an iterative build-up is also possible in the context of our DQMC approach if needed.


\section{The Hubbard chain}
\label{sec:Hubbard}
To demonstrate the applicability and numerical efficiency of our replica switching method to calculate \Renyi entropies within the DQMC framework we study the entanglement of a one-dimensional Hubbard chain, which at half-filling does not suffer from the infamous sign problem.
We first concentrate on zero-temperature properties of the entanglement entropy. We compare our numerical results to the quasi-exact analytical form derived from the conformal field theory description of the gapless theory governing the zero-temperature physics of the Hubbard chain in the presence of a finite on-site interaction $U$. 
We then turn to finite-temperature properties and show how the \Renyi entropy  crosses over from a low-temperature entanglement entropy to a high-temperature thermal entropy. 
Finally, we discuss the strength of our approach in comparison to a recent proposal to calculate \Renyi entropies for interacting fermion systems from a decomposition in free fermion Green's functions \cite{grover_entanglement_2013}. We demonstrate that our approach is significantly more efficient in capturing the entanglement properties in the interaction dominated regime of the Hubbard model.

\subsection{Zero-temperature physics}
In the presence of a repulsive on-site interaction $U>0$ the ground state of the half-filled Hubbard chain is well known to be a Mott insulator exhibiting quasi-long range antiferromagnetic order. This means that at zero temperature charge fluctuations are frozen out entirely for any $U>0$, while the localized spin degrees of freedom interact via an effective Heisenberg exchange of order $t^2/U$ thereby building up quasi-long range antiferromagnetic order. The system thus remains gapless and can be described in terms of a conformal field theory with a central charge $c=1$ corresponding to the number of gapless modes.

\begin{figure}\label{fig:gs_oscillations}
\includegraphics{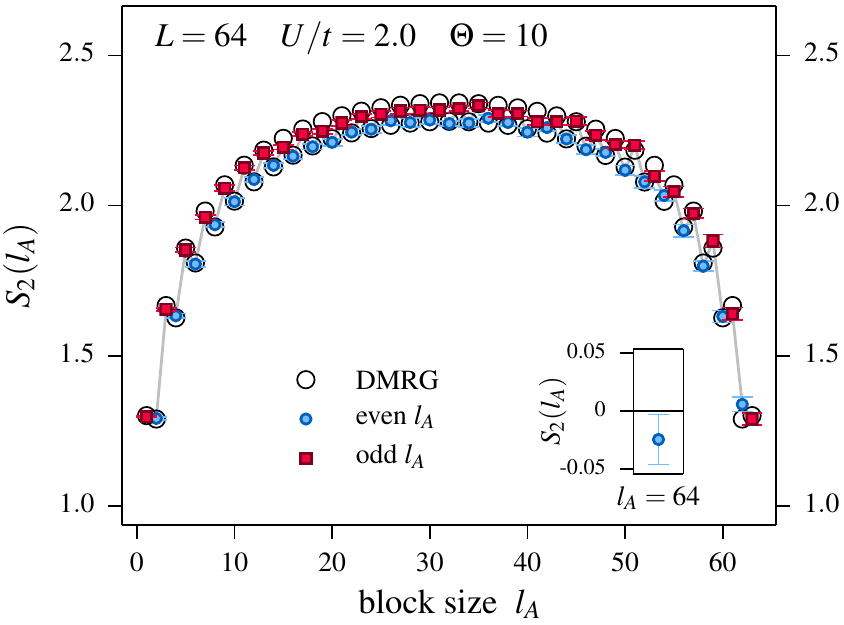}
  \caption{(Color online)  The entanglement entropy $S_2$ of a periodic, half-filled Hubbard chain with $L=64$ sites in the presence of a 	
  		repulsive on-site
                 interaction $U/t=2$. Shown is the entanglement entropy versus the length of the subsystem $l_A$. 
                 The numerical data obtained with the zero-temperature DQMC algorithm $(\Theta=10)$ is in 
                 good agreement with DMRG reference data for the same system (open circles).
                  The inset shows $S_2(L)$, i.e. the entanglement entropy for a subsystem 		equal to the whole chain. A value of $0$ 
                  corresponds to a perfect purity  
                  $\mc{P} = 1$ of the sampled ground state and thus signifies that the 	
                  projection parameter $\Theta$ in the zero-temperature DQMC approach 
                  was chosen sufficiently large. 
\label{fig:central_charge}}
\end{figure}
The entanglement entropy of such a gapless one-dimensional system is known \cite{Holzhey,VidalKitaev,CalabreseCardy} to exhibit a logarithmic correction to the boundary law, which for a one-dimensional system simply states that the entanglement entropy is a constant for any bipartition. The logarithmic correction, however, does reflect the relative size of the two subsystems in the bipartition and for all \Renyi entropies follows the 
general form \cite{CalabreseCardy}
\begin{equation}
  S_n(l_A) = \dfrac{c}{6\eta} \left( 1+\frac{1}{n}\right) \ln{ \left[ \dfrac{\eta L}{\pi} \sin{\left( \dfrac{\pi l_A}{L}\right)} \right]} +O(1)  \,,
  \label{eq:cft_of_l}
\end{equation}
where $c$ is the central charge of the conformal field theory, $L$ is the overall system length and $l_A\leq L$ is the length of subsystem $A$.
Open and periodic boundary conditions correspond to $\eta=2$ and $\eta=1$, respectively, and further subleading corrections of order $O(1)$ in the system size are neglected.
Numerical results obtained with the zero-temperature DQMC algorithm (for details see the appendix) for an open chain of length $L = 64$ 
are shown in Fig.~\ref{fig:central_charge}. We find that the DQMC data is generally in good agreement with quasi-exact results obtained using density matrix renormalization group (DMRG) simulations. We do observe, however, a slight trend of the DQMC data to underestimate (within the statistical error bars) the entanglement entropy for some of the intermediate block sizes -- an effect which we find to be absent for smaller system sizes (not shown) and which previously has also been observed in conceptually similar simulations for spin systems~\cite{hastings_measuring_2010} using the replica trick in combination with a stochastic series expansion (SSE) \cite{Footnote}. 

We thus conclude this section with the observation that our replica switching DQMC method is indeed well equipped to efficiently determine the zero-temperature entanglement properties of the half-filled Hubbard chain.


\subsection{Thermal crossover of the entanglement}

\begin{figure}
\includegraphics[width=\columnwidth]{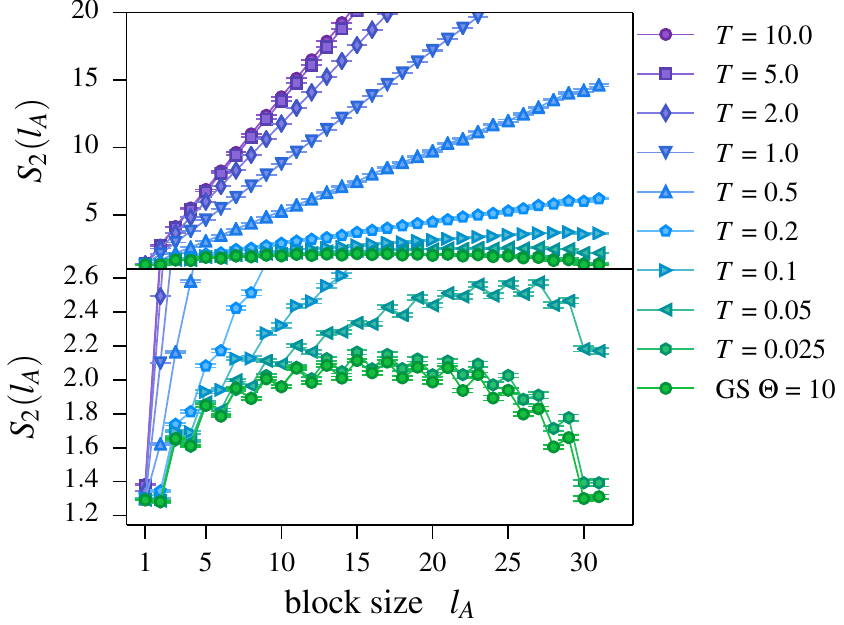}
\caption{(color online) The thermal crossover of the entanglement entropy for a $L=32$ site half-filled Hubbard chain with $U/t = 2$. 
	      While at high temperatures the entanglement entropy exhibits a volume law $S=\log{4}\cdot L$ (indicated by the dashed line),
	      it crosses over to a boundary law at low temperatures with a characteristic arc-like structure.
	      \label{fig:oscillations_crossover}}
\end{figure}

When considering a quantum system at {\em finite} temperatures, both quantum and thermal fluctuations contribute to all entropies
including the \Renyi entropy of interest here. Increasing the temperature the relative contributions of these two types of 
fluctuations of course change. As a result the \Renyi entropy shows a crossover from a boundary law (with logarithmic corrections)
at zero temperature to a more conventional extensive behavior (i.e. a volume law) at high temperature of the form
\[
	S(l_A) = l_A \cdot \log 4 \,,
\]
simply counting the number of possible states in the subsystem.

\begin{figure}
\includegraphics[width=\columnwidth]{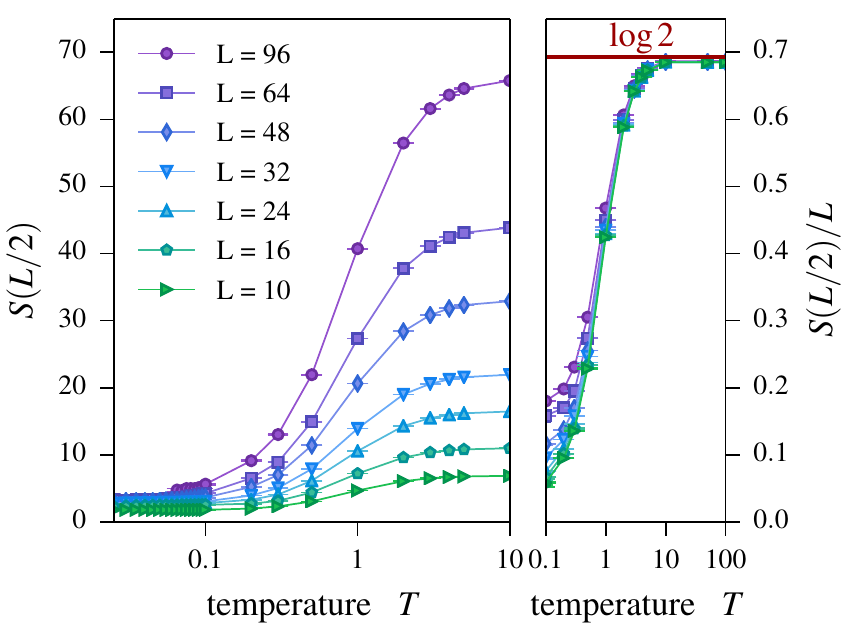}
\caption{(color online) Thermal crossover of the \Renyi entropy for a half-filled Hubbard chain with $U/t=1$. 
	      Shown is $S_2(L/2,T)$ (left panel) and a rescaled $S_2(L/2,T)/L$ (right panel), which at high temperatures
	      converges to $\log 2$ for different system sizes. 
	      \label{fig:s2_half_chain}}
\end{figure}

This thermal crossover of the \Renyi entropy from a zero-temperature entanglement entropy to a thermal entropy at high temperatures can easily be observed in our numerical DQMC simulations. 
This is illustrated in Fig.~\ref{fig:oscillations_crossover} for a half-filled Hubbard chain of length $L=32$ with intermediate on-site interaction $U/t=2$ in a temperature range $0.025 \leq T \leq 5$ (for $t=1$). With increasing temperature the arc-like structure of the low-temperature entanglement entropy disappears and gives way to the simple linear form of an extensive thermal entropy.
This thermal crossover is also reflected in Fig.~\ref{fig:s2_half_chain} where we plot the \Renyi entropy $S_2(L/2)$ of an equal-size bipartition of the chain for different system sizes versus temperature. In particular, we observe the expected data collapse at high temperatures when rescaling the calculated \Renyi entropies by the respective system size, see the right panel of Fig.~\ref{fig:s2_half_chain}.

\begin{figure}
\includegraphics[width=\columnwidth]{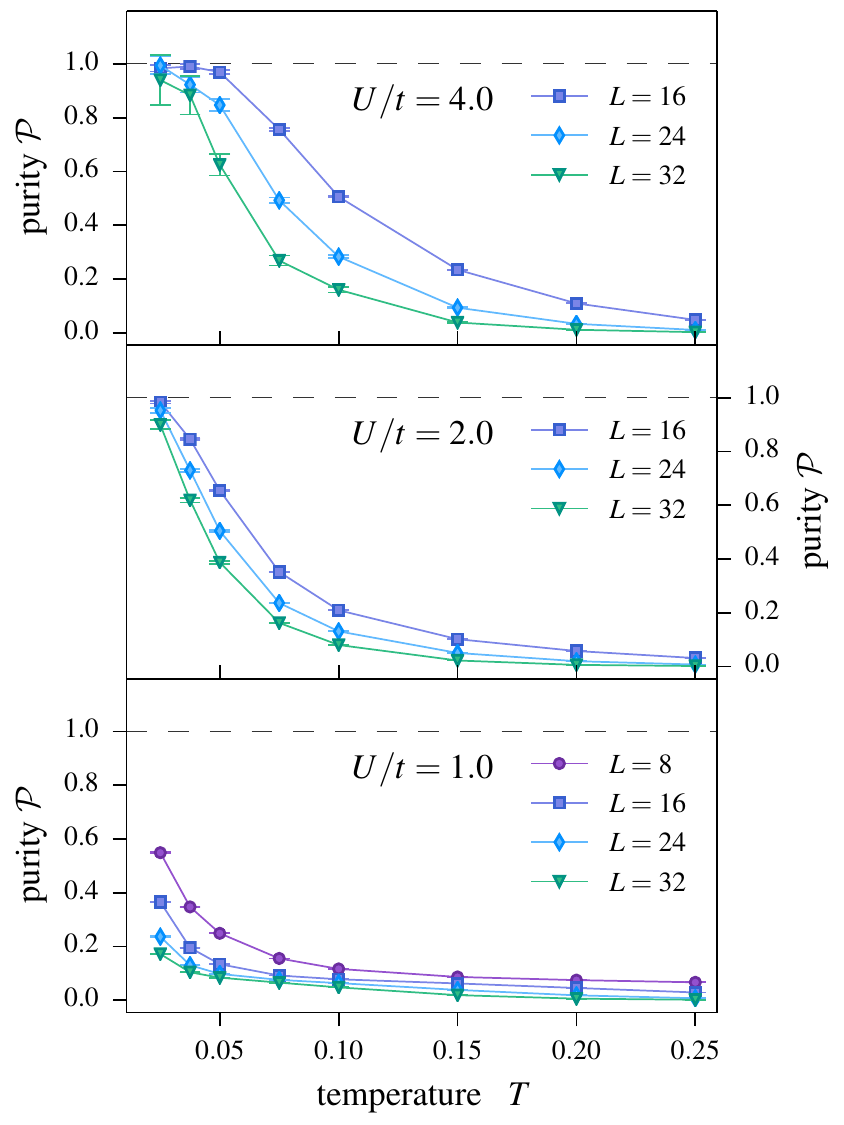}
\caption{(color online) The purity $\mc{P}$ for a grand-canonical DMQC simulation of a half-filled Hubbard chain versus temperature for varying on-site interactions $U/t$ and  chains of varying length $L$. 
}\label{fig:purity}
\end{figure}

To quantitatively determine the crossover temperature $T^*$, below which a finite-sized system is effectively in its ground state, 
we turn to the so-called {\em purity}
\begin{equation}
	\mc{P} = \exp{\left(-S_2(L) \right)} \,,
\end{equation}
which becomes $1$ for a quantum mechanical ground state, since the entropy $S_2(L)$ needs to equal its complement $S_2(\emptyset)$
and thus must vanish for any quantum mechanical ground state -- an observation which is ultimately also responsible for the arc-like structure of the entanglement entropy in Eq.~\eqref{eq:S2}. Indeed we find that the purity sharply rises towards $1$ as the temperature is lowered in our simulations, see Fig.~\ref{fig:purity} where we plot the purity as a function of temperature for different system sizes and a sequence of on-site interactions. On the one hand, we find that for a fixed value of the on-site interaction the crossover temperature decreases with system size in accordance with the fact that the finite-size gap of the system also decreases with increasing system size. On the other hand, we observe that for fixed system size the transition temperature $T^*$ decreases as the on-site interactions $U$ is reduced reflecting the enhancement of charge fluctuations in this weakly coupled regime. 

Finally, we note that we generally find somewhat smaller transition temperatures than recent stochastic series expansion (SSE) simulations~\cite{bonnes_entropy_2013} of the half-filled Hubbard chain, which can be tracked back to the fact that our DQMC simulations employ a {\em grand-canonical} ensemble, while the SSE simulation in Ref.~\onlinecite{bonnes_entropy_2013} employed a canonical ensemble.

\subsection{Comparison to free fermion decomposition method}

\begin{figure}
\includegraphics[width=\columnwidth]{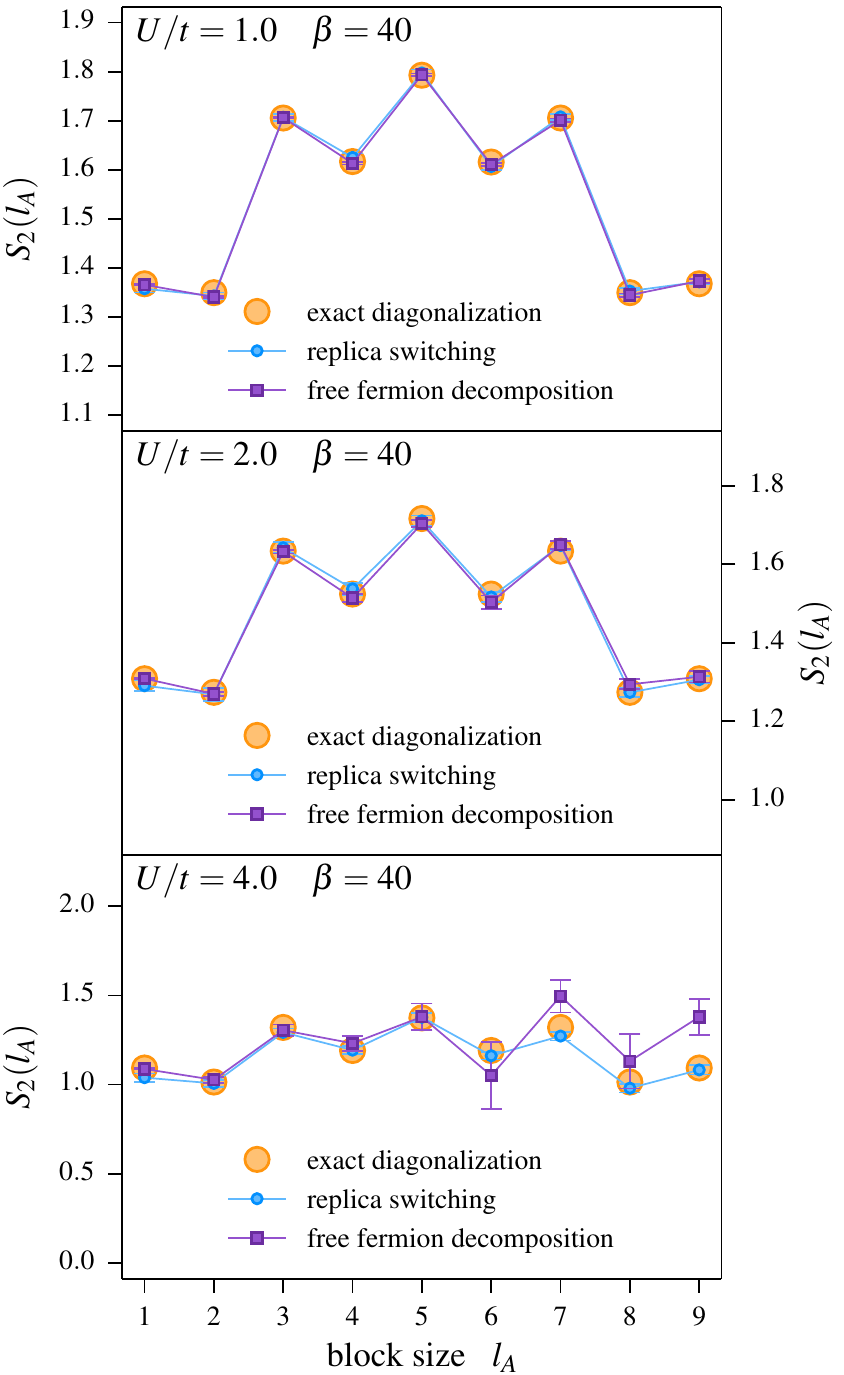}
\caption{(color online) Comparison of the replica switching (squares) and free fermion decomposition (circles) DQMC algorithms for the \Renyi entropy of a half-filled Hubbard chain with varying on-site interactions $U/t$ at temperature $T = 0.025$ $(\beta=40)$. For comparison exact diagonalization data (for $\beta=\infty$) is indicated by the open circles. With both codes allocating the exact same amount of CPU time, a much faster convergence of the replica switching method is found.}
\label{fig:comparison}
\end{figure}

We round off our discussion of our replica switching DQMC method by comparing it to a recent proposal~\cite{grover_entanglement_2013} 
to calculate  \Renyi entropies from a decomposition in free fermion Green's functions. Such a decomposition might seem natural within the DQMC approach, since the Hubbard-Stratonovich transformation at its heart results in an effective description of free fermions moving in an external potential.

In the spirit of a fair comparison we have implemented the free fermion decomposition method~\cite{grover_entanglement_2013} using the same algorithmic optimizations as for our replica switching technique whenever possible. We then ran both codes on identical parameter sets logging the exact same CPU time for both codes. Results from this comparison are summarized in Fig.~\ref{fig:comparison} where we show
results of both approaches for the entanglement entropy of a half-filled 10-site Hubbard chain at fixed temperature $T=0.025$ $(\beta=40)$ for different values of the on-site interaction $U/t \in \{1, 2, 4\}$. While the free fermion decomposition method reproduces the arc-like structure of the entanglement entropy for small on-site interaction $U/t=1$, it shows deviations from this behavior already for moderate values of the on-site interaction $U/t=2$. In contrast, our replica switching method nicely reproduces the exact diagonalization data up to strong on-site interaction $U/t=4$.
We thus conclude that our approach is significantly more efficient in capturing the entanglement properties in the interaction dominated regime of the Hubbard model.


\section{Conclusions}
\label{sec:outlook}

In summary, we have introduced a novel replica switching scheme to efficiently calculate \Renyi entropies for interacting fermionic systems in determinantal quantum Monte Carlo simulations. Our approach is capable of efficiently determine not only finite-temperature thermal entropies but also zero-temperature entanglement entropies as demonstrated for the half-filled Hubbard chain over a range of interactions. In comparison to a recent proposal to calculate \Renyi entropies from a decomposition in terms of free fermion Green's functions \cite{grover_entanglement_2013}, our approach yields much faster convergence and significantly higher numerical efficiency in the regime of strong correlations $U/t > 1$.

While we have concentrated our discussion on the second \Renyi entropy $S_{n=2}$, it should be noted that the replica technique described here can be expanded in a straight-forward way to also access higher \Renyi entropies. Calculating \Renyi entropies with higher indices $n>2$, however,  requires to simulate a system of size $N_A + n\cdot N_B$ at temperature $n \beta$ (where $N_A$ and $N_B$ correspond to the size of subsystems $A$ and $B$, respectively). Thus, the computational cost to access these higher \Renyi entropies in our approach sharply increases as the DQMC simulations generally scale as $\beta N^3$ (where $N$ is the total number of sites). This should be contrasted to the free-fermion decomposition method of Grover \cite{grover_entanglement_2013}, where the \mbox{$n$-th} \Renyi entropy can readily be accessed by simply simulating $n$ replicas of the system at temperature $\beta$ -- a much more moderate increase in computational expense. Indeed recent numerical simulations  \cite{assaad_entanglement_2013} for the Kane-Mele model have demonstrated that with the free fermion decomposition method it is feasible to calculate a partial entanglement spectrum, i.e. the spectrum of (lowest) eigenvalues of the density matrix, from a sequence of \Renyi entropies. Likely, such a calculation of the entanglement spectrum is out of reach for our technique. 

Although we have focused our discussion on one-dimensional fermion systems in this manuscript, two dimensional systems can be treated equally well. In particular,
we point out that the numerical overhead of our method arising from the simulations of the two copies $B$ and $B'$ (see Fig.~\ref{fig:EnsembleSwitchingUnfolded}) for the second \Renyi entropy reduces with increasing spatial dimensionality: In one dimension an equal-size bipartition results in an overhead of $L/2$ additional sites simulated at temperature $2 \beta$ . In two spatial dimensions, where we might consider a subsystem $A$ of size $L/2 \times L/2$ and its somewhat larger complement, we can choose to simulate the smaller subsystem (in this case $A$) twice. Such a simulation would thus only need an overhead of $1/4$ of the sites in comparison to a conventional simulation and thus make it possible to calculate \Renyi entropies for systems of nearly the same size as conventionally investigated in DQMC simulations.

\subsection*{Acknowledgments}
We thank F. Assaad, T. Grover, and L. Bonnes for discussions. We also thank M. Becker for providing the DMRG reference data of Fig.~\ref{fig:central_charge} and T. Grover for providing the exact diagonalization data of Fig.~\ref{fig:comparison}.
We acknowledge support from SFB TR 12 and the Bonn-Cologne Graduate School of Physics and Astronomy. 
The numerical simulations were performed on the CHEOPS cluster at RRZK Cologne.


\appendix 
\label{sec:supplemental}

\section{DQMC Primer}
\label{app:DQMC}
We will give a short introduction to the basics of determinantal Quantum Monte Carlo that should suffice to understand the modififications to it presented in the main text.
There are many more extensive, excellent reviews on the DQMC method available, such as Refs.~\onlinecite{santos_introduction, assaad_world-line} among others, on which our exposition is based.
We will start with the finite temperature algorithm and later mention what modifications are necessary to perform ground state simulations. For concreteness, we will stay within the context of the Hubbard model.

\subsection{Finite-Temperature Algorithm}
The first step in setting up the algorithm is to Trotter decompose imaginary time:
\begin{align}
 \mc{Z} &= \tr{\prod\limits_i \exp{\left(- \Delta\tau  \mc{H} \right)}} \\
&= \sum\limits_{\{\ket{\psi}\}} \bra{\psi}\exp{\left(- \Delta\tau  \mc{H} \right)} \dots  \exp{\left(- \Delta\tau  \mc{H} \right)}\ket{\psi}. \label{eq:discretizedpi}
\end{align}
The occupation number operator $\mc{N}$, needed for the chemical potential, is from now on included in the potential operator $\mc{V}$. 
The exponential is now separated and the appearing commutator ignored which in turn results in a systematic error of the order $\mc{O}\left(\Delta\tau\right)$.
\[ e^{A + B} \approx e^A e^B \quad\Rightarrow\quad e^{-\Delta\tau (\mc{K} + \mc{V})} \approx e^{-\Delta\tau \mc{K}} e^{-\Delta\tau \mc{V}} + \mc{O}\left(\Delta\tau\right)\]
The kinetic part $e^{-\dtau \mc{K}}$ has only two operators and can be evaluated directly. 
The potential part $e^{-\dtau \mc{V}}$ on the other hand is made up of four operators and can therefore not be integrated out analytically.
We thus apply the Hubbard-Stratonovich transformation to decouple the interaction. It is based on an identity for Gaussian integrals
\begin{equation}\label{hstrafo}
e^{\frac{1}{2}A^2} = \sqrt{2\pi} \infint{\meas{}{s} e^{-\frac{1}{2} s^2 - sA}}
\end{equation}
with $A$ being the original operator, in our case the on-site interaction $\mc{V}_i = U n_{i, \uparrow} n_{i, \downarrow}$. 
The price we have to pay is the introduction of the eponymous auxiliary field $s$ that couples to the fermions.
The decoupling itself is not unique: we can decouple with respect to the local charge or to the local magnetization.
For the repulsive case, we usually couple to the local magnetization to avoid a complex transformation which is nevertheless possible and can be advantageous~\cite{assaad_su2-spin_1998}. We do, however, break $SU(2)$ symmetry which is only restored when performing enough updates.

It turns out that in the case of the Hubbard model, it is not necessary to work with a continuous auxiliary field, 
but that we can choose to work with discrete Ising spins $s$ taking values $\{ -1, 1 \}$~\cite{hirsch_discrete_1983}. 
The decoupling has to be performed for each site and at each time slice. Doing so, we obtain
\begin{align}
e^{-U\dtau n_{i, \uparrow} n_{i, \downarrow}} &= \dfrac{1}{2} e^{-\onehalf \dtau U n_i}\sum\limits_{s = \pm 1}{e^{-\lambda s m_i}}\nonumber \\
&= \donehalf\sum\limits_{s = \pm 1}\prod\limits_{\sigma = \uparrow, \downarrow}{e^{-\left(\sigma s \lambda + \onehalf U\dtau\right)n_{i, \sigma}}}\nonumber
\end{align}
The parameter $\lambda$ can be determined by inserting all possible values for the auxiliary field $s = \pm 1$ and the spins $\sigma = \{ \uparrow , \downarrow\}$. We find
\[\cosh{\lambda} = e^{\onehalf \abs{U}\dtau}\]
Inserting this into \eqref{eq:discretizedpi} and switching to an explicit vector notation for the operators, we obtain the following form for the partition function:
\[ \mc{Z} = \left(\donehalf\right)^{N^dL} \text{Tr}_{\{s\}} \tr{} \prod\limits_{l = L}^1 \prod\limits_{\sigma = \uparrow, \downarrow} e^{\dtau \tbf{c}_\sigma^\dagger \tbf{K} \tbf{c}_\sigma} e^{-\dtau \tbf{c}^\dagger_\sigma \tbf{V}^\sigma_s (l) \tbf{c}_\sigma},\]
where $\tbf{K}$ and $\tbf{V}$ are the matrix representations of the $\mc{K}$ and $\mc{V}$ operator, respectively.

The partition function now consists of one trace over all auxiliary field configurations and another one over the fermionic states which act on a product over all time slices and spins of the discretized and separated exponential.

We continue to rewrite the operators in matrix form $\tbf{K}$ and $\tbf{V}^\sigma_s$ where the index $s$ reminds us that the potential part depends on the auxiliary field.
A product $e^{\dtau \tbf{c}_\sigma^\dagger \tbf{K} \tbf{c}_\sigma} e^{-\dtau \tbf{c}^\dagger_\sigma \tbf{V}_s^\sigma (l) \tbf{c}_\sigma}$ will be denoted by $\tbf{B}^\sigma (l)$
and an ordered sequence of all $L$ $\tbf{B}$-matrices with arbitrary starting point $\tau$ by
\begin{equation}\label{eq:b_mat_sequence}
\mc{B}^\sigma(\tau) = \tbf{B}^\sigma (\tau - 1)\dots \tbf{B}^\sigma(0)\tbf{B}^\sigma(\beta)\dots\tbf{B}^\sigma(\tau).
\end{equation}
The time $\tau$ is a multiple of the discretization $\dtau$. The weight of one configuration of fermions and auxiliary field is then given by
\begin{equation}\label{eq:fermionweight}
\bra{\psi}\mc{B}^\sigma(\tau)\ket{\psi} = \det\left(\mc{B}^\sigma(\tau)\right) \,.
\end{equation}
We could use this weight to set up the Monte Carlo simulation, sampling both fermion and auxiliary field configurations. But this would be very inefficient because the calculation of determinants is numerically very expensive. Instead, we will integrate out the fermions analytically and sample only auxiliary field configurations.

One can prove that the \textit{grand-canonical} trace over all fermionic states of the $\tbf{B}$-matrix is given by 
\begin{equation}\label{eq:det_ident}
\tr{\,\mc{B}} = \det{(1 + \mc{B})},
\end{equation}
 allowing us to explicitly perform the trace over all fermionic states.
One can hardly underestimate the importance of this identity for determinantal QMC simulations. By applying the identity~\eqref{eq:det_ident}, we are left only with the problem of sampling the auxiliary field where the weight of each configuration is given by one determinant. The connection to the field is made by the $\mc{B}$ matrices via the potential. 
 Using these notations, we can rewrite the partition function as
\[\mc{Z} = \left(\donehalf\right)^{N^dL}\sum\limits_{\{ s\}}\prod\limits_\sigma \det{(1 + \mc{B}^\sigma(0))}.\]
In principle, we are now able to set up our Monte Carlo simulations. We know the form of the weights and we can sample configurations of auxiliary fields using, for example, the Metropolis scheme mentioned before. For the calculation of Green's functions and more practical aspects we refer the interested reader to the aforementioned references.

\subsection{Ground-State Algorithm}
\label{app:GSalgorithm}
To access ground state properties, we start with a carefully chosen trial wavefunction $\ket{\psi_T}$ and project out the excited states by applying a large power of the Hamiltonian:
\begin{equation}\label{eq:projection}
\lim\limits_{\Theta \rightarrow \infty} e^{-\Theta\mc{H}}\ket{\psi_T} = \ket{\psi}
\end{equation}
The trial wavefunction must be non-orthogonal to the true ground state wavefunction $\ket{\psi}$ for the procedure to work. For the Hubbard model at half filling we chose a spin singlet as trial wave function  $\ket{\psi_T}$.
Setting up the simulation is very similar to the finite temperature algorithm. 
The projection parameter $\Theta$ plays the role of the temperature $\beta$.
The exponential in~\eqref{eq:projection} is then decomposed and the interaction term decoupled by a Hubbard-Stratonovich transformation. 
We will not go into further detail because the calculation of the \Renyi entropies for ground state problems was shown to reduce to a modified finite temperature problem.
\subsection{Numerical Optimizations}
In DQMC simulations we are limited by the $N^3\beta$ scaling and have to pay special attention to numerical instabilities. 
Simulating an artificially enlarged system may thus appear to be very inefficient.
However, enlarging the system and implementing the Hamiltonian~\eqref{eq:img_time_ham} goes along with a special matrix structure that we can make use of to lower the numerical cost.
In the matrix representation, the hopping terms are of the following form:
\begin{align}
\tilde{\tbf{K}}_{AB} &= 
\left(\begin{matrix} 
  \tbf{K}_{AA} & \tbf{K}_{AB} & 0 \\
  \tbf{K}_{BA} & \tbf{K}_{BB} & 0 \\
  0 & 0 & \mathbbm{1} 
\end{matrix}\right)
,\;
\tilde{\tbf{K}}_{AB^\prime} = 
\left(\begin{matrix} 
  \tbf{K}_{AA} & 0 & \tbf{K}_{AB^\prime} \\
  0 & \mathbbm{1} & 0 \\
  \tbf{K}_{B^\prime A} & 0 & \tbf{K}_{B^\prime B^\prime}\nonumber
\end{matrix}\right),
\end{align}
where $\tbf{K}$ is the matrix representation of the hopping operator and the index signifies the part of the lattice it connects. One of the parts $B$ and $B^\prime$, respecticely, remains unchanged and thus has a $\mathbbm{1}$ on the diagonal and zeros for the off-diagonal terms. Interaction terms will be of the same form. 

\begin{figure}
\vspace{0.2cm}
\centering
\includegraphics[width=\columnwidth]{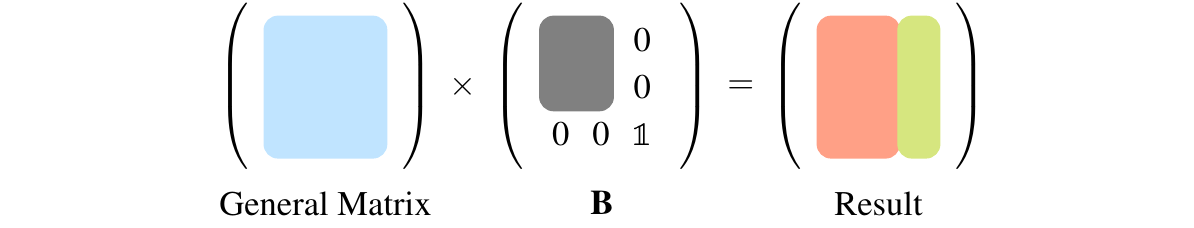}
\caption{(color online) The computational cost of the algorithm due to the additional sites is lowered by exploiting the matrix structure induced by the imaginary time dependent Hamiltonian~\eqref{eq:img_time_ham}. The slice matrices comprise blocks of zeros which should explicitely be ignored in steps involving matrix multiplications.}\label{fig:matrix_multiplication}
\end{figure}

One can now readily convince oneself that multiplying two matrices in the same imaginary time interval $(0, \beta)$ or $(\beta, 2\beta)$ will not alter this structure because they are block-diagonal. Mixing matrices of different time intervals, on the other hand, will typically result in a dense matrices. This is inevitably the case when calculating the sequence of $\tbf{B}$ matrices~\eqref{eq:b_mat_sequence}, except for $\tau \in \{0, \beta\}$. 
Even when multiplying a dense matrix with one of the slice matrices, it is not necessary to perform a full matrix multiplication, because of the blocks of zeros present in the slice matrix.

\begin{figure*}
\includegraphics[width=2\columnwidth]{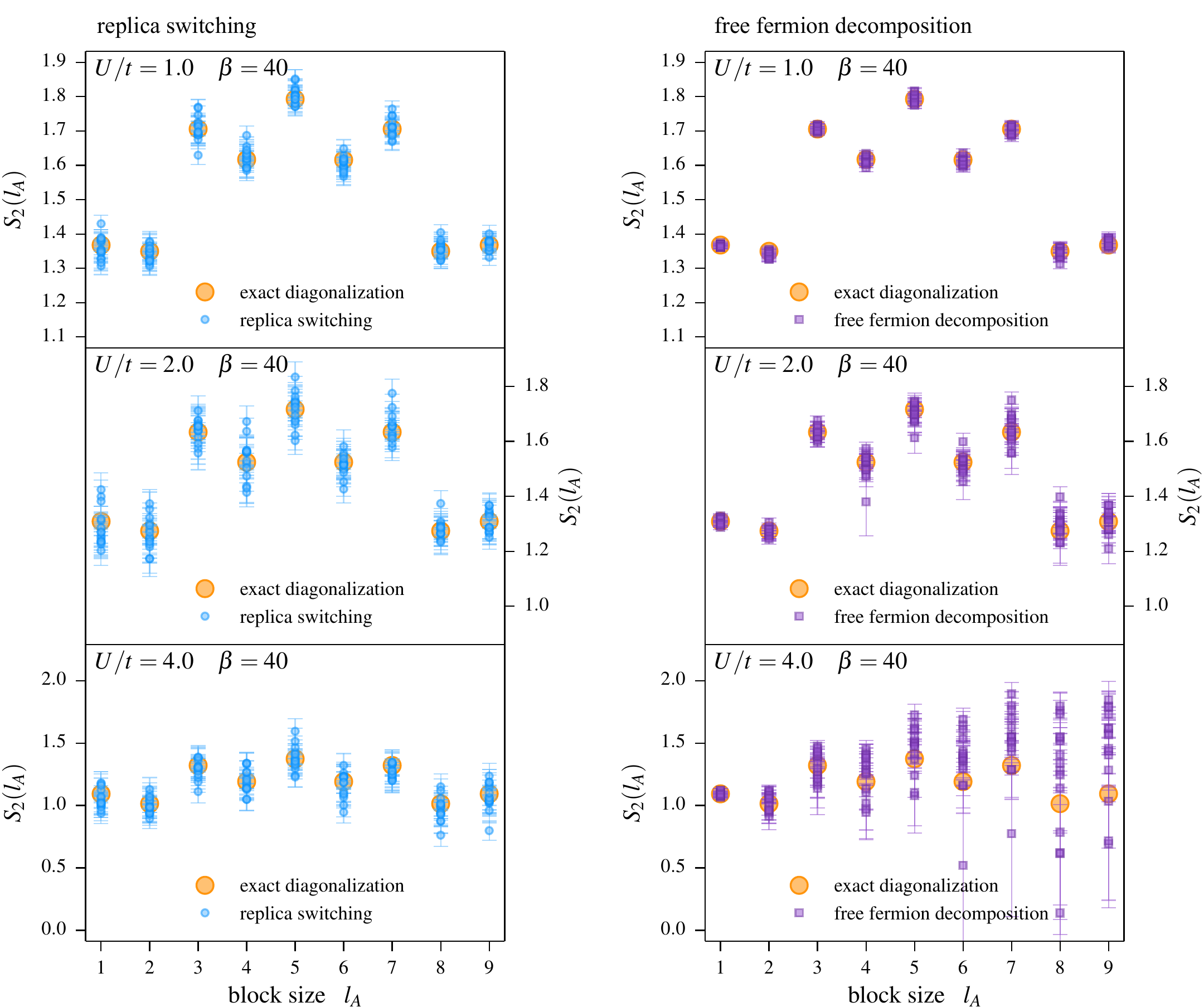}
\caption{(color online) Comparison of the spread of the raw data of DQMC runs for 16 independent runs using the  
 the replica switching (squares) and free fermion decomposition (circles)  algorithms. 
Shown is the \Renyi entropy of a half-filled Hubbard chain with varying on-site interactions $U/t$ at temperature $T = 0.025$ $(\beta=40)$.}
\label{fig:unmerged_comparison}
\end{figure*}

\subsection{The ensemble switching method}
\label{app:ensemble_switching}
We will show that to estimate the ratio~\eqref{eq:S2} it suffices to measure the expectation values of $\expv{p_{1\rightarrow 2}}$ and $\expv{p_{2\rightarrow 1}}$, which are the probabilities to switch ensembles from ensemble $1$ to $2$ and from ensemble $2$ to $1$, respectively.
Using a Metropolis scheme, the probability to switch ensembles for a given configuration $c$ is given as
\begin{equation}
 p_{1\rightarrow 2} = \min{\left(1, \dfrac{W_2(c)}{W_1(c)}\right)}\, ,\nonumber
\end{equation}
where $W_1(c)$ and $W_2(c)$ are the weights in the respective ensembles. The configurations $c$ are configurations of the auxiliary field and both ensembles share the same configuration space $\mc{C}$. 
Writing out the ratio of the expectation values for the switching operation gives
\begin{align}
\dfrac{\expv{p_{1\rightarrow 2}}}{\expv{p_{2\rightarrow 1}}} &= 
\dfrac{\mc{Z}_2}{\mc{Z}_1} \cdot \dfrac{\sum\limits_{c\,\in\,\mc{C}(\mc{Z}_1)} \min{\left(1, \dfrac{W_2(c)}{W_1(c)}\right)}\, W_1(c)}{\sum\limits_{c\,\in\,\mc{C}(\mc{Z}_2)} \min{\left(1, \dfrac{W_1(c)}{W_2(c)}\right)}\, W_2(c)}\nonumber\\
&= \dfrac{\mc{Z}_2}{\mc{Z}_1} \cdot \dfrac{\sum\limits_{W_2(c) < W_1(c)} W_2(c) + \sum\limits_{W_1(c) < W_2(c)} W_1(c)}{\sum\limits_{W_1(c) < W_2(c)} W_1(c) + \sum\limits_{W_2(c) < W_1(c)} W_2(c)}\nonumber\\
&= \dfrac{\mc{Z}_2}{\mc{Z}_1}\, .\nonumber
\end{align}

\section{Comparison to free fermion decomposition}

In this appendix we provide further details on our analysis of the comparison between the replica switching and free fermion decomposition DQMC algorithms as presented in Fig.~\ref{fig:comparison} of the main text. 

Our {\rm C++} simulation codes of the two algorithms employ a common code base implementing the same optimizations for many underlying core features (such as linear algebra subroutines, sampling improvements or parallelization schemes) for both approaches and further build on the ALPS libraries~\cite{alps}. 

We separately ran 16 independent simulations per data point in Fig.~\ref{fig:comparison} of the main text each with a different random seed and later merged the results via a jackknife analysis using the ALPS tools. The spread of the raw results of all 16 runs before merging are shown in Fig.~\ref{fig:unmerged_comparison}. 

For an allocated computing time of 105 minutes (per data point and seed) the number of measurements in the free fermion case  was around 3500 per data point and seed (after initial thermalization). For the replica switching method, we obtained a considerably smaller number of measurements for the switching probabilities in~\eqref{eq:switching_ratio_main} in the same allocated computing time.
For the largest cut, i.e. $l_A = 10$, we recorded some 1300 measurements, while for the smallest cut, i.e. $l_A = 1$ we recorded only some 100 measurements.
The lower number of measurements for the ensemble switching method in a given time frame is due to two effects: First, for the replica switching method we have to perform one simulation for an ensemble of two separate systems and one simulation in the connected ensemble. Both of these simulations have to be thermalized in contrast to only one simulation in the free fermion case. Second, the simulation cell of the connected system in the replica switching method is enlarged. For the case of $l_A = 1$ for example, we effectively simulate a system of size $L = 19$ which in combination with the $N^3\beta$ scaling of the algorithm further reduces the number of possible sweeps in a given time.

Comparing the results obtained with the two algorithms the spread of the raw data shows two overall trends. First, looking at the dependence of the spread of data points with subsystem size $l_A$, the free fermion decomposition method shows a considerable increase of this spread with increasing subsystem size, while for the replica switching method there is no measurable dependence. Second,  with increasing onsite interaction $U$ the data spread clearly increases much stronger for the free fermion decomposition method in comparison with the replica switching technique (which becomes most poignant for large subsystem sizes $l_A$).



\begin{thebibliography}{99}
\bibitem{Footnote}
In the context of DQMC simulations, one additional potential origin for the systematic bias might be the discretization of imaginary time. 
In our simulations it was generally set to a value of $\Delta\tau = 0.125$, which for the studied onsite interaction of $U = 2.0$ is smaller than typically needed~\cite{santos_introduction}.
\end{thebibliography}
\end{document}